\documentclass[cup5b]{caps}
\usepackage{graphicx}
\usepackage{amssymb}
\usepackage{ociwsymp2}
\HeadText{J. H. Schwarz}

\begin{document}

\pagenumbering{arabic}

\author[]{JOHN H. SCHWARZ\\ California Institute of Technology}

\chapter{Update on String Theory}

\begin{abstract}

The first part of this report gives a very quick sketch of how
string theory concepts originated and evolved during its first 25
years (1968-93). The second part presents a somewhat more detailed
discussion of the highlights of the past decade. The final part
discusses some of the major problems that remain to be solved.

\end{abstract}

\section{Introduction}

There are two primary goals in fundamental physics. The first is
to construct a unified theory, incorporating relativity and
quantum theory, that describes all fundamental forces, as well as
all of the elementary particles.  The second goal is to understand
the origin and evolution of the Universe. Needless to say, these
are both incredibly ambitious objectives, representing the
extremes of microphysics and macrophysics. The only reason that we
can even pose them with a straight face is the fact that so much
progress toward each of these objectives has already been
achieved.

Superstring theory (also known as M-theory) is a promising
candidate for the fundamental underlying theory. It used to be
believed that superstring theory is a collection of theories, but
as we will discuss, it now appears that (in a certain precise
sense) there is a unique theory with no adjustable parameters. But
this theory is a work in progress --- not yet fully formulated ---
and it is unclear how far its elucidation will take us toward
realizing these lofty goals. Even in the most optimistic scenario,
it will certainly take a long time.

This talk will take a somewhat historical tack in describing the
subject. The first part will give a very quick sketch of how
string theory concepts originated and evolved during its first 25
years (1968-93). In the second part a somewhat more detailed
discussion of the highlights of the past decade will be presented.
The final section will list and comment on some of the major
problems that remain to be solved. The reader who wants a more
thorough treatment is referred to the standard texts (Green,
Schwarz, \& Witten 1987; Polchinski 1998)

\section{1968--1993}

String theory arose in the late 1960s in an attempt to construct a
theory of the strong  nuclear force. Experiments in the 1960s
revealed a rich spectrum of strongly interacting particles
(hadrons) lying on nearly straight and parallel Regge
trajectories. This means that there were families of particles
identified whose spin increased linearly with the square of their
mass. This point of view was developed as part of the S matrix
theory/bootstrap program that was fashionable in the 1960s. The
linear Regge trajectories were interpreted in terms of poles of
the analytic S matrix in the angular momentum plane. This picture
successfully accounted for certain asymptotic properties of
scattering amplitudes at high energy. This led Veneziano (1968) to
propose a very simple mathematical function (basically the Euler
beta function) as an approximate expression for a scattering
amplitude that realized these properties. Over the next two years
a small community of theorists generalized this to formulas for
$N$-body scattering amplitudes with various interrelated consistency
properties. The fact that this could be done suggested the
possibility that Veneziano's formula was not just a
phenomenological amplitude (as originally intended) but actually
part of a full-fledged theory.

Soon thereafter it was recognized independently by Nambu (1970),
Susskind (1970), and Nielsen (1970) that the theory that was being
developed could be understood physically as one based on
one-dimensional structures (called strings), rather than
pointlike particles. It then became clear how such a theory could
account for various qualitative features of hadrons and their
interactions. The basic idea is that specific particles correspond
to specific oscillation modes of the string. S matrix theory, the
bootstrap program, and Regge poles were abandoned (for good
reasons) a long time ago. However, one cannot deny that they have
left a lasting legacy. String theory probably could have been
discovered by another route, but it might have taken many more
years for that to happen. (Witten used to say that string theory
is 21st century physics that happened to be discovered in the 20th
century.) Be that as it may, the attempts to construct a string
theory of hadrons were not fully successful. Moreover, in the
early 1970s a quantum field theory description of the strong force
--- namely quantum chromodynamics (QCD) --- was developed. It was
universally accepted and string theory fell out of favor.

\subsection{Problems with String Theory}

Even though string theory had many attractive qualitative
features, it began to unravel when one demanded that it should
provide a full-fledged self-consistent theory of the hadrons. The
original version of string theory turned out to have several fatal
flaws. One was that consistency of perturbation theory, beyond the
tree approximation, requires 25 spatial dimensions and one time
dimension. A second major shortcoming is that this theory does not
contain any fermions. Moreover the perturbative spectrum contains
tachyons and massless particles. The former imply an unstable
vacuum, and the latter are not part of the hadron spectrum.

\subsection{Supersymmetry}

In an attempt to do better,  a string theory that does contain
fermions was introduced (Neveu \& Schwarz 1971; Ramond 1971). It
requires only nine spatial dimensions --- not the desired answer,
but a step in the right direction. The original version of this
theory also had a tachyon in its spectrum, but it was later
realized that there is a consistent truncation that eliminates it.
Very importantly, the study of this theory led to the modern
understanding of supersymmetry and superstrings. In the original
version (with the tachyon) supersymmetry was present only on the
two-dimensional world-sheet, a fact that was first pointed out by
Gervais \& Sakita (1971). The pioneering work of Wess \& Zumino
(1974) in the construction of supersymmetric quantum field
theories was motivated by the search for four-dimensional analogs
of this two-dimensional symmetry. A few years later, Gliozzi,
Scherk, \& Olive (1976) realized that the truncation that
eliminates the tachyon also results in 10-dimensional spacetime
supersymmetry.

Supersymmetry is an important idea for several reasons. For one
thing, given very weak assumptions, it is the unique possibility
for a nontrivial extension of the usual (Poincar\'e group)
symmetries of space and time. Moreover, it is a symmetry that
relates bosons and fermions. That is, they belong to common
irreducible representations. Most importantly, there is a good
chance that the new particles required by supersymmetry will be
discovered at accelerators in this decade. Basically every known
elementary particle that is a fermion (i.e., the quarks and
leptons) should have bosonic partners and every known elementary
particle that is a boson (the gauge bosons and Higgs) should have
fermionic partners.

There are several different reasons to expect that the masses of
the superpartners should be roughly at the TeV scale. Finding them
(perhaps at the LHC) may provide the best experimental support for
superstring theory in the foreseeable future. There is a good
chance that the lightest super particle (the LSP) is absolutely
stable. It is a leading candidate for a cold dark matter WIMP. So
it is possible that supersymmetry will be discovered first in dark
matter searches.

\subsection{Gravity and Unification}

One of the major obstacles to using these string theories to
describe hadron physics was the occurrence of massless particles
in the string spectrum. This conflicts with the fact that all
hadrons are massive. We spent several years unsuccessfully trying
to construct string theories that look more like realistic hadron
theories. That did not work, yet string theory (especially the
10-dimensional one) was too beautiful to abandon. So we finally
decided to try to understand it on its own terms.

Eventually, it was realized (Scherk \& Schwarz 1974; Yoneya 1974)
that one of the massless particles in the closed string spectrum
is spin 2 and interacts at low energy in exactly the right way to
be identified as a graviton --- the quantum of gravitation.
Therefore Scherk and I (1974) proposed using string theory to
describe gravity and unification rather than just the strong
force. This requires the typical size of a string to be about
$10^{-32}$ cm (the Planck scale) rather than $10^{-13}$ cm (the
typical size of a hadron), as was previously assumed.

This proposal had two immediate benefits: (1) Previous approaches
to quantum gravity give unacceptable infinities. String theory
does not. (2) Extra dimensions can be acceptable in a gravitational
theory, where spacetime geometry is determined dynamically. For
these reasons, as well as the fact that string theory requires
gravity, Scherk and I were convinced that superstring unification
was an important idea. However, for the next 10 years only a few
brave souls shared our enthusiasm.

\subsection{The First Superstring Revolution}

One of those brave souls was Michael Green. In 1984 my five-year
collaboration with Green culminated in an explanation of how
superstring theory can be compatible with parity violation (Green
\& Schwarz 1984), which is an important feature of the standard
model. (This was previously considered to be impossible.)
Following that other superstring theories were found (Gross et al.
1985), and there were specific proposals
for the geometry of the extra six dimensions (Candelas et al.
1985), which come surprisingly close to
explaining the standard model.

By the time the dust settled we had five consistent superstring
theories:
\[ {\rm I, ~~IIA, ~~IIB, ~~HE, ~~HO,}\]
\noindent each of which requires 10 dimensions. The Type I and HO
theories each have an SO(32) gauge group, whereas the HE theory
has an $E_8 \times E_8$ gauge group. Each of these theories has
${\cal N} = 1$ supersymmetry in the 10-dimensional sense. This is
the same amount of supersymmetry as  what is called ${\cal N} = 4$
in four dimensions (16 conserved supercharges). The two Type II
theories have twice as much supersymmetry, and do not exhibit any
nonabelian gauge symmetry in 10 dimensions. It took another 10
years to figure out how to achieve nonabelian gauge symmetry
(using D-branes) in these theories.

The most realistic schemes (in those days) involved the HE theory
with six dimensions forming a Calabi-Yau space. A Calabi-Yau
space is a K\"ahler manifold of SU(3) holonomy. Such manifolds
have been much studied, since their relevance to string theory was
recognized, and by now there is a great deal known about them.
They have two important topological integers associated with them,
called $h_{11}$ and $h_{21}$. $h_{11}$ is the dimension of the
space of K\"ahler forms, and $h_{21}$ is the dimension of the space
of complex structure deformations. In the context of the
compactification of the HE theory, one ends up at low energy with
a four-dimensional gauge theory with ${\cal N} = 1$ supersymmetry
and gauge group $E_6$, though if the Calabi-Yau space is not
simply connected this can be broken through the addition of Wilson
lines to a gauge group very close to that of the standard model,
possibly with additional U(1) factors. The number of generations
of quarks and leptons is given by $|h_{11} - h_{21}|/2$. There are
a number of known Calabi-Yau spaces that give the desired answer
(three), but there are thousands of others that give different
answers. There is no particular mathematical reason to prefer any
of the three generation models. In any case, the analysis is
carried out for weak coupling, and it is not clear that this is a
good approximation.

One also needs to analyze the couplings of the gauge theory. Some
relevant information, in the case of models with ${\cal N} = 1$
supersymmetry, is encoded in a holomorphic function called the
superpotential. One wants to know the superpotential not only in
the classical approximation but also for the quantum effective
action. Over the years one has learned how to derive the effective
superpotential, at least in certain cases, and to read off whether
interesting phenomena such as confinement or dynamical
supersymmetry breaking are implied by its structure.

With these discoveries string theory became a very active (though
somewhat controversial) branch of theoretical high-energy physics.
Over the subsequent 18 years it has become even more active (but
less controversial).

\subsection{T Duality}

String theory exhibits many strange and surprising properties. In
fact, one could argue that it represents a conceptual revolution
comparable to that associated with quantum theory. One surprising
property that was discovered in the late 1980s is called T
duality. (The letter T has no particular significance. It was the
symbol used by some authors for one of the low-energy fields.) A
good  review has been written by Giveon, Porrati, \& Rabinovici
(1994). When it is relevant, T duality implies that two different
geometries for the extra dimensions are physically equivalent! For
example, a circle of radius $R$ can be equivalent to a circle of
radius $\ell^2/R$, where $\ell$ is the fundamental string length
scale. (String theorists often use $\alpha' = \ell^2$, which is
the Regge slope parameter.)

Let me sketch an argument that should make this duality plausible,
since at first sight it is highly counterintuitive. When there is
a circular extra dimension, the momentum along that direction is
quantized: $p=n/R$, where $n$ is an integer. Using the
relativistic energy formula $E^2 = M^2 + \sum_i (p_i)^2$ (in units
with $c=1$), one sees that the momentum along the circular
dimension can be interpreted as contributing an amount $(n/R)^2$
to the mass squared as measured by an observer in the noncompact
dimensions. This is true whether one is considering point
particles, strings, or any other kinds of objects. However, in the
special case of closed strings, there is a second kind of
excitation that can also contribute to the mass squared. Namely,
the string can be wound around the circle, so that it is caught up
on the topology of the space. The string tension is (now also
setting $\hbar =1$) given by the string scale as $1/(2\pi
\ell^2)$. The contribution to the mass squared is the square of
this tension times the length of wrapped string, which is $2\pi R
m$, if it wraps $m$ times. Multiplying, the contribution to the
mass squared is $(R m /\ell^2)^2$. Now we can make the key
observation: under T duality the role of momentum excitations and
winding-mode excitations are interchanged. Note that the
contributions to the mass squared match if one interchanges $m$
and $n$ at the same time that $R \to \ell^2/R$.

Usually T duality relates two different theories. Two
particularly important examples are
\[ {\rm IIA ~~\leftrightarrow ~~IIB} \quad {\rm and} \quad
{\rm HE ~~\leftrightarrow ~~HO}.\] Therefore IIA \& IIB (also HE
\& HO) should be regarded as a single theory. More precisely, they
represent opposite ends of a continuum of geometries as one varies
the radius of a circular dimension. This radius is not a parameter
of the underlying theory. Rather, it arises as the vacuum
expectation value of a scalar field. Thus, in principle it is
determined dynamically, though in the most symmetrical examples
there is a flat potential so that any value is possible. (Such
scalar fields that do not appear in the potential are called
``moduli.'' Moduli probably should not be part of a realistic
solution, since they tend to give a scalar component to long-range
gravitational strength forces.)

These T duality identifications reduce the list from five to three
superstring theories. When other equivalences (such as S duality
discussed below) are also taken into account, we conclude that
there is actually a unique theory! What could be better? We are
led to a unique underlying theory, free from arbitrary parameters,
as the only possibility for a consistent quantum theory containing
gravity. Well, there are a few details that still need to be
fleshed out. (See the final section of this report.)

There are also fancier examples of T-duality-like equivalences,
such as the physical equivalence of Type IIA superstring theory
compactified on a Calabi-Yau space and Type IIB compactified on
the ``mirror" Calabi-Yau space. This mirror pairing of
topologically distinct Calabi-Yau spaces was a striking
mathematical discovery made by physicists (Greene \& Plesser
1990), which has subsequently been explored by mathematicians. The
two Hodge numbers $h_{11}$ and $h_{21}$, discussed above, are
interchanged in the mirror transformation.

T duality suggests a possible way for a big crunch to turn into a
big bang (Brandenberger \& Vafa 1989). The heuristic idea is that
a contracting space when it becomes smaller than the string scale
can be reinterpreted as an expanding space that is larger than the
string scale, without the need for any exotic forces to halt the
contraction. However, to be perfectly honest, we do not yet have
the tools to analyze such time-dependent scenarios reliably.

T duality implies that usual geometric concepts break down at the
string scale. Another manifestation of this is ``noncommutative
geometry,'' which arises when certain fields are turned on
(Connes, Douglas, \& Schwarz 1998; Seiberg \& Witten 1999). It
results in an ``uncertainty relation'' of the form $\Delta x
\Delta y \geq \theta$, analogous to the more familiar $\Delta x
\Delta p \geq h$, limiting one's ability to localize a particle in
two orthogonal directions at once.

\section{1994--Present}

For its first 25 years string theory was studied entirely in terms
of perturbation expansions in a string coupling constant $g$,
which characterizes the strength of interaction. The Feynman
diagrams of string theory are given by two-dimensional surfaces
that represent the string world sheet. The story is especially
simple in the case of oriented closed strings, since then there is
a single Feynman diagram at each order of the perturbation
expansion, in striking contrast to quantum field theory. The
classification is given in terms of closed Riemann surfaces, with
the genus corresponding to the number of loops. Perturbation
theory is the way one often studies quantum field theory, as well.
As is known from the example of QED, this can work very well when
the coupling is small. However, as illustrated by QCD at low
energy, for example, such perturbation expansions are not the
whole story. New phenomena (such as confinement and chiral
symmetry breaking in the case of QCD) arise when $g$ is not small.

In string theory $g$ is not a free parameter. Rather, it is
determined dynamically as the value of a certain string field (the
dilaton). When perturbation theory makes sense the dilaton is one
of the moduli. In a realistic model it probably should not  be a
modulus, and then other methods of calculation may be required.

\subsection{S Duality}

Another kind of duality --- called S duality --- was discovered as
part of the ``second superstring revolution'' in the mid 1990s. S
duality allows us to go beyond perturbation theory. It relates $g$
to $1/g$ in the same way that T duality relates $R$ to $1/R$. The
two basic examples (Hull \& Townsend 1995; Witten 1995) are
\[ {\rm I ~~\leftrightarrow ~~HO} \quad {\rm and} \quad
{\rm IIB ~~\leftrightarrow ~~IIB}.\]
Thus we learn how these three
theories behave when $g \gg 1$. For example, strongly coupled Type
I theory is equivalent to the weakly coupled SO(32) heterotic
theory.

The transformation $g$ to $1/g$ (or, more precisely, the
corresponding transformation of the dilaton field) is a symmetry
of the Type IIB theory. In fact, this is a subgroup of an infinite
discrete symmetry group SL(2, Z). If some of the extra dimensions
are compactified to give a torus, even larger discrete symmetry
groups that combine the S duality and T duality groups arise (Hull
\& Townsend 1995), in which case one sometimes speaks of U duality.
This story has been reviewed by Obers \& Pioline (1999).

\subsection{M Theory}

S duality tells us how three of the five original superstring
theories behave at strong coupling. This raises the following
question: what happens to the other two superstring theories ---
IIA and HE --- when $g$ is large? The answer is quite remarkable:
They grow an 11th dimension of size $g\ell$ . This new
dimension is a circle in the IIA case (Townsend 1995; Witten 1995)
and a line interval in the HE case (Horava \& Witten 1995). It is
not visible in perturbation theory, since that involves expansions
about $g=0$.

When the 11th dimension is large, one is outside the regime of
perturbative string theory, and new techniques are required. At
low energies, the 11-dimensional theory can be approximated by
11-dimensional supergravity (Cremmer, Julia, \& Scherk 1978).
However, that is only a classical theory. What we need is a
full-fledged quantum theory, for which Witten has proposed the
name M theory. Since it is unclear how best to formulate it, he
imagined that M stands for ``mysterious'' or ``magical.'' Others have
suggested ``membrane'' and ``mother.''

One more or less obvious idea is to try to construct a realistic
four-dimensional theory by starting in 11 dimensions and
choosing a suitable 7-manifold for the extra dimensions. This
raises the question: what 7-manifolds are suitable? It is
generally assumed that one wants to end up with a low-energy
theory that looks more or less like the minimal supersymmetric
extension of the standard model, which has ${\cal N} =1$
supersymmetry. That is what one achieved by Calabi-Yau
compactification of the HE theory.

Starting from M theory, one can prove that the way to get  ${\cal
N} =1$ supersymmetry is to require that the 7-manifold have $G_2$
holonomy. $G_2$ is one of the exceptional Lie groups --- the only
one that can occur as a holonomy group. The study of $G_2$
manifolds is more difficult and less well understood than that of
Calabi-Yau manifolds. Relatively few examples are known. One
basic fact is that if the $G_2$ manifold is smooth, the resulting
four-dimensional theory cannot have nonabelian gauge symmetry or
chiral fermions. Therefore to get interesting models, it is
necessary to consider $G_2$ manifolds with particular kinds of
singularities. This has been an active area of study in the past
few years, and some progress has been made, but there is still a
long way to go. It is possible that some models constructed this
way will turn out to be dual to ones constructed by Calabi-Yau
compactification of the HE theory. That would be interesting,
because each description would be better suited for exploring
certain regimes. For example, the M theory picture should allow
one to understand phenomena that are nonperturbative in the
heterotic picture.

\subsection{p-branes}

Superstring theory requires new objects, called $p$-branes, in
addition to the fundamental strings. ($p$ is the number of spatial
dimensions; e.g., a string is a 1-brane.) All $p$-branes, other
than the fundamental string, become infinitely heavy as $g \to 0$,
and therefore they do not appear in perturbation theory. On the
other hand, at strong coupling this distinction no longer applies,
and they are just as important as the fundamental strings.

Superstring theory contains a number of higher rank analogs of
gauge fields with antisymmetrized indices, which are interpreted
geometrically as differential forms. They can couple to higher
dimensional objects, much like U(1) gauge fields can couple to
charged point particles. Specifically, a $(p+1)$-form gauge field
can couple electrically to a $p$-brane or magnetically to a
$(d-p-3)$-brane, where $d$ is the number of spacetime dimensions.
Because of supersymmetry, the energy density of such a charged
brane is bounded from below, and when the (BPS) bound is
saturated, a certain amount of the supersymmetry remains unbroken,
and the brane is stable.

\subsection{D-branes}

An important class of $p$-branes, called D-branes, has the
(defining) property that fundamental strings can end on them
(Polchinski 1995). This implies that quantum field theories are
associated with D-branes. These field theories are of the
Yang-Mills type, like the standard model (Witten 1996). D-branes
have a tension (or energy density) that scales with the string
coupling like $1/g$. This is to be contrasted with the more
characteristic solitonic $1/g^2$ behavior, exhibited by the
NS5-branes, which are the magnetic duals of the fundamental
strings in the Type II and heterotic theories.

An interesting possibility is that we experience four dimensions,
because we are confined to live on D3-branes, which are embedded
in a spacetime with six additional spatial directions. To be
compatible with the observed $1/r^2$ force law for gravity, the
extra dimensions would need to either have a size $\ll 1$ mm (so
as not to have been detected in Cavendish-type experiments) or
else be very ``warped'' --- which means that the 4d geometry
depends on the position in the other six dimensions.

The stable D-branes in the IIA theory have an even number of
spatial dimensions, whereas those in the IIB theory have an odd
number of spatial dimensions. Thus, as a particular example, the
IIA theory contains D0-branes, which are pointlike objects. These
are like extremal black holes that carry just one unit of a
certain conserved $U(1)$ gauge symmetry charge. Recall that the
IIA theory actually has a circular 11th dimension. The $U(1)$
charge is nothing but the integer that characterizes the momentum
along this circle. Thus a D0-brane actually has one unit of
momentum along the circle. As we have argued earlier, the mass of
such a particle (if there is no other contribution to its mass)
should just be $1/R$. However, we have said that the radius is
$g\ell$. Therefore, one deduces that the mass of a D0-brane must
be $1/(g\ell)$, which is in fact the correct value.

As we indicated earlier, 11-dimensional M theory requires a
precise quantum definition, which is not provided by
11-dimensional supergravity. A specific proposal, called
matrix theory, was put forward by Banks et al.
(1997). In this proposal one builds the theory out of $N$
D0-branes and considers a limit in which $N \to \infty$. This
corresponds to going to the infinite momentum frame along the
compactified circle. In the limit one can also decompactify the
circle, so as to end up with a description of M theory in a flat
11-dimensional spacetime. This proposal has passed a number of
nontrivial tests, and it is probably correct. However, it is
awkward to work with. In particular, 11-dimensional Lorentz
invariance seems very mysterious in this setup.

\subsection{Black Hole Entropy}

The gravitational field of the D-branes causes warpage of the
spacetime geometry and creates event horizons, like those
associated with black holes.  In fact, studies of D-branes have
led to a much deeper understanding of black hole thermodynamics in
terms of string theory microphysics. In special cases, starting
with a five-dimensional example analyzed by Strominger \& Vafa
(1996), one can count the microstates associated with D-brane
excitations and compare then with the area of the corresponding
black hole's event horizon, confirming the existence of
statistical physics underpinnings for the Bekenstein-Hawking
entropy formula. Although many examples have been studied and no
discrepancies have been found, we are not yet able to establish
this correspondence in full generality. The problem is that one
needs to extrapolate from the weakly coupled D-brane picture to
the strongly coupled black hole one, and mathematical control of
this extrapolation is only straightforward when there is a
generous measure of unbroken supersymmetry.

\subsection{AdS/CFT}

In a remarkable development, Maldacena (1997) conjectured that the
quantum field theory that lives on a collection of D3-branes (in
the IIB theory) is actually equivalent to Type IIB string theory
in the geometry that the gravitational field of the D3-branes
creates. This proposal was further elucidated by Witten (1998) and
by Gubser, Klebanov, \& Polyakov (1998). An excellent review
followed a couple years later (Aharony et al.
2000).

In the IIB/D3-brane example a certain (conformally invariant)
four-dimensional quantum field theory (CFT), called  ${\cal N} =
4$ super Yang-Mills theory (Brink, Scherk, \& Schwarz 1977;
Gliozzi, Scherk, \& Olive 1977), is precisely equivalent to Type
IIB string theory in a 10-dimensional spacetime that is a product
of a five-dimensional anti de Sitter (AdS) spacetime and a
five-dimensional sphere (Schwarz 1983). Maldacena also proposed
several other analogous dualities. This astonishing proposal has
been extended and generalized in a couple thousand subsequent
papers.

While I cannot hope to convince you here that these AdS/CFT
dualities are sensible, I can point out that the first check is
that the symmetries match. An important ingredient in this
matching is the fact that the isometry group of anti de Sitter
space in $D+1$ dimensions is SO(D,2), the same as the conformal
group in $D$ dimensions. The fact that the four-dimensional gauge
theory with ${\cal N} = 4$ supersymmetry is conformal at the
quantum level is itself a highly nontrivial fact discovered around
1980. It requires that all the UV divergences cancel and the
coupling constant does not vary with energy (vanishing beta
function) so that renormalization does not introduce a scale.

The study of AdS/CFT duality and its generalizations is serving as
a theoretical laboratory for exploring many deep truths about the
inner works of superstring theory and its relation to more
conventional quantum field theories. As one example of the type of
insight that is emerging, let me note that by breaking the
conformal symmetry and deforming the AdS geometry one can
construct examples in which renormalization group flow in the
four-dimensional quantum field theory corresponds to radial motion
in the higher dimensional string theory spacetime.

This correspondence raises the hope that (among other things) we
can close the circle of ideas --- and solve our original problem,
namely to find a string theory description of QCD, the theory of
the strong nuclear force! That problem is now viewed as a low-energy
analog of the unification problem. As yet, the
10-dimensional geometry that gives the string theory dual of QCD
has not been identified. There are a number of technical hurdles
that need to be overcome that I will not go into here. Presumably,
none of them is insuperable.

On the other hand, the duality (discussed above) between  supersymmetrical
Yang-Mills
theory with  ${\cal N} = 4$ supersymmetry and SU(N) gauge symmetry,
and Type IIB superstring theory in an $AdS_5
\times S^5$ spacetime background with $N$ units of five-form
(Ramond-Ramond) flux, are very beautiful and surely correct.
However, our inability to carry out concrete string theory
calculations in this background has been a source of frustration.
Even though this geometry has almost as much symmetry as flat
spacetime, calculations are much more difficult. As a result,
most studies use a low-energy supergravity approximation to the
string theory, which corresponds to restricting the dual gauge
theory to a certain corner of its parameter space, where gauge
theory calculations are difficult. Despite a number of interesting
suggestions, I think it is fair to say that no practical scheme
for doing string theory calculations in this background is known.
While this problem has not  been solved, there is an interesting
limit in which it can be sidestepped.

\subsection{The Plane-wave Limit}

Recently, using a method due to Penrose (1976), Blau et al.
(2002) constructed a
plane-wave limit of the Type IIB superstring in the $AdS_5 \times
S^5$ spacetime background.  The limiting plane-wave background is
also maximally supersymmetric. Following that, Metsaev (2002)
showed that Type IIB superstrings in this plane-wave background
are described in the Green-Schwarz light-cone gauge formalism by
free massive bosons and fermions, so that explicit string
calculations are tractable. The essential difference from flat
space is the addition of mass terms in the world sheet action,
something that would have been regarded as very peculiar without
this motivation. Then Berenstein, Maldacena, \& Nastase (2002)
identified the corresponding limit of the gauge theory and carried
out some checks of the duality. I will describe this subject in a
little more detail than any of the preceding ones, because it is
my current research interest.

Let me first describe how the Penrose limit works in this case.
The AdS space (in global coordinates) is described by the metric
\[ds^2(AdS_{5}) = R^2 (-\cosh^2 \rho\, dt^2
+ d\rho^2 + \sinh^2 \rho \, d \Omega_3^2),\] where the scale $R$
has been factored out. Similarly the five-sphere is described by
the metric
\[ds^2(S^{5}) = R^2 (\cos^2 \theta \,d\phi^2
+ d\theta^2 + \sin^2 \theta \, d \tilde{\Omega}_3^2).\] In order
to describe the desired limit, it is convenient to make the
changes of variables
\[r = R \sinh \rho, \quad y = R \sin \theta\]
\[x^+ = t/\mu, \quad x^- = \mu R^2 (\phi - t). \]
Here $\mu$ is an arbitrary mass scale. The coordinate $x^-$ has
period $2\pi \mu R^2$, since $\phi$ is an angle.  Therefore the
conjugate (angular) momentum is
\[ P_- = J/\mu R^2,\]
where $J$ is an integer.

In terms of the new coordinates, the $AdS_5 \times S^5$ metric
becomes
\[ ds^2 = 2 \left(1 - {y^2}/{R^2}\right)
dx^+ dx^- -\mu^2 (r^2 + y^2) (dx^+)^2\]
\[+ \frac{1}{\mu^2 R^2} \left( 1 - {y^2}/{R^2}
\right) (d x^-)^2 + ds^2_\perp, \]
where
\[ ds_\perp^2 = r^2 d\Omega_3^2 + \frac{R^2}{R^2 + r^2} dr^2
 + y^2 d\tilde{\Omega}_3^2 + \frac{R^2}{R^2 - y^2} dy^2\]
and \[ y^2 = \sum_{I = 1}^{4} (x^I)^2 ~ {\rm and} ~  r^2 = \sum_{I
= 5}^{8} (x^I)^2.\] The limit $R \rightarrow \infty$ gives the
desired plane-wave geometry:
\[ds^2 = 2d x^+ dx^- - \mu^2 (x^I)^2 (dx^+)^2 + dx^I dx^I.\]
This differs from flat 10-dimensional Minkowski spacetime only by
the presence of the mass term, which acts rather like a confining
harmonic oscillator potential in the eight transverse dimensions.

This subject has aroused a great deal of interest over the past
year, because for the first time we have a setup in which
tractable gauge theory calculations can be compared to tractable
string theory calculations. However, neither of these is easy, and
there are any number of subtle issues to be sorted out. So this
keeps a lot of clever people busy.

Let me sketch some of the essential features of the duality.
Calling the coupling constant of the gauge theory $g_{YM}$,
$\lambda = g_{YM}^2 N$ is a combination that 't Hooft identified
long ago as important in the large-$N$ limit. In the AdS/CFT
duality $g_{YM}^2$ corresponds to the string coupling $g$, and
$\lambda$ corresponds to $(R/\ell)^4$ in the string theory. The
plane-wave limit introduces some additional identifications. For
one thing the gauge theory has a global SU(4) symmetry. Picking an
arbitrary U(1) subgroup and calling the associated charge $J$, one
identifies $J$ with  $\mu R^2 P_-$ in the plane-wave geometry. (We
argued earlier that this should be an integer.) Furthermore the
limit $R \rightarrow \infty$ corresponds in the gauge theory to
letting $ J, N \rightarrow \infty$ while holding fixed the
combination $\lambda' = {g_{YM}^2 N}/{J^2}$, which corresponds to
$(\ell^2 \mu P_-)^{-2}$. The gauge theory expansion parameter
$\lambda'$ for correlation functions of the Berenstein et al.
(BMN) operators replaces the usual $\lambda = g_{YM}^2 N$. This is
possible because the BMN operators are almost supersymmetric in
the large-$N$ limit. This may be a lot to swallow if you have not
seen it before. But hopefully, you get the general idea.

To explore this duality in detail, it is necessary to establish a
precise dictionary between single-trace gauge-invariant BMN
operators and string states. This is fairly straightforward to
leading order in the string coupling, but at higher order the
correct matching requires mixing of single-trace and double-trace
operators in the gauge theory. For the BMN sector of the gauge
theory, the perturbation expansion can be organized as a double
expansion in $\lambda'$ (instead of $\lambda$) and $g_2 = J^2/N$.
The order in the latter parameter corresponds to the  genus of the
Feynman diagram (when expressed in double line notation following
't Hooft). In the string side of the duality, the perturbation
expansion corresponds to the expansion in $g_2$, but each order
contains the complete $\lambda'$ dependence, both perturbative and
nonperturbative. Therefore computations of the first couple of
orders in the string side provide powerful predictions for the
gauge theory.

The leading order (in $g_2$) string prediction was given in the
original BMN paper. It has been verified to all orders in
$\lambda'$ in the dual gauge theory (Santambrogio \& Zanon 2002).
The information at this order consists of a comparison of
anomalous dimensions of BMN operators in the gauge theory and
light-cone energies in the string theory. The next order in $g_2$
involves comparing three-string couplings with more complicated
correlators in the gauge theory, and is much more challenging.

The only way known to formulate the order $g_2$ interactions in
the string theory is in terms of light-cone gauge string field
theory in the plane-wave geometry. The formulas in flat space were
worked out long ago (Green \& Schwarz 1983; Green, Schwarz, \&
Brink 1983). However, in flat space there are more efficient
formalisms, so that the light-cone gauge string field theory
approach was largely ignored until this year when the need arose
to generalize it to the plane-wave geometry. That has been
achieved by Spradlin \& Volovich (2002, 2003) and by
Pankiewicz \& Stefanski (2003). However, the formulas involve
inverses of infinite matrices, which need to be computed before
one can make explicit gauge theory predictions to all orders in
$\lambda'$. I am pleased to report that this has been achieved
within the last couple of weeks (He et al.  2002).

\section{Some Remaining Problems}

Let me conclude by listing, and briefly commenting upon, some of
the issues that still need to be resolved, if we are to achieve
the lofty goals indicated at the beginning of this report. As will
be evident, most of these are quite daunting, and the solution of
any one of them would be an important achievement. Let me begin
with the list that a particle theorist might make.

\begin{itemize}
\item {\it Find a complete and optimal formulation of the theory.}
Although we have techniques for identifying large classes of
consistent quantum vacua, we do not have a succinct and compelling
formulation of the underlying theory of which they are vacua. Many
things that we take for granted, such as the existence of a
spacetime manifold, should probably be emergent properties of
specific vacua rather than identifiable features of the underlying
theory. If this is correct, then we clearly need something that is
quite unlike all theories with which we are familiar. It is
possible that when the proper formulation is found the name
``string theory'' will become obsolete.

\item {\it Understand why the cosmological constant (the energy
density of empty space) vanishes.} The exact value may be nonzero
on cosmological scales, but a Lorentz invariant Minkowski
spacetime, which requires a vanishing vacuum energy, is surely an
excellent approximation to the real world for particle physics
purposes. We can achieve an exact cancellation between the
contributions of bosons and fermions when there is unbroken
supersymmetry. There does not seem to be a good reason for such a
cancellation when supersymmetry is broken, however. Many
imaginative proposals have been made to solve this problem, and I
have not studied each and every one of them. Still, I think it is
fair to say that none of them has gained a wide following. My
suspicion is that the right idea is yet to be found.

\item {\it Find all static solutions (or quantum vacua) of the
theory.} This is a tall order. Very many families of consistent
supersymmetric vacua, often with a large number of moduli, have
been found. The analysis becomes more difficult as the amount of
unbroken supersymmetry decreases and the moduli (including the
dilaton) are eliminated or made massive. Vacua without
supersymmetry are a real problem. In addition to the issue of the
cosmological constant, one must also address the issue of quantum
stability. Stable nonsupersymmetric classical solutions are often
destabilized by quantum corrections. As far as I am aware, there
are no known examples for which this has been proved not to
happen.

\item {\it Determine which quantum vacuum describes all of
particle physics and understand whether it is special or just an
environmental accident.} Presumably, if one had a complete answer
to the preceding item, one of the quantum vacua would be an
excellent approximation to the microscopic world of particle
physics. Obviously, It would be great to know the right solution,
but we would also like to understand why it is the right solution.
Is it picked out by some special mathematical property or is it
just an accident of our particular corner of the Universe? The way
this plays out will be important in determining the extent to
which the observed world of particle physics can be deduced from
first principles.
\end{itemize}

Next let me turn to the list that a cosmologist might make.

\begin{itemize}

\item {\it Understand why the cosmological constant (the energy
density of empty space) is incredibly small, but not zero.} As you
know, current observational evidence suggest that about 70\% of
the closure density is provided by negative pressure ``dark
energy.'' The most straightforward candidate is a small
cosmological constant, though other possibilities are being
considered. Whether or not a cosmological constant is the right
answer, string theorists would certainly be pleased if they could
give a compelling reason why it should vanish. (See the second
item on the particle physics list.) We would then be in a better
position to study possible sources of tiny deviations from zero.

\item {\it Understand how string theory prevents quantum
information from being destroyed by black holes.} Long ago,
Hawking (1976) suggested that when matter falls into black holes
and eventually comes back out as thermal radiation, quantum
coherence is lost. In short, an initial pure state can evolve into
a mixed state, in violation of the basic tenets of quantum
mechanics. I am convinced that string theory is a unitary quantum
theory in which this can never happen, and so Hawking must be
wrong. Still, as far as I know, nobody has formulated a complete
explanation of how string theory keeps track of quantum phase
relations as black holes come and go.

\item {\it Understand when and how string theory resolves
spacetime singularities.} Singularities are a generic feature of
nontrivial solutions to general relativity. Not only are they
places where the theory breaks down, but, even worse, they undermine
the Cauchy problem --- the ability to deduce the future from
initial data. The situation in string theory is surely better.
Strings sense spacetime differently than point particles do.
Certain classes of spacelike singularities, which would not be
sensible in general relativity, are known to be entirely harmless
in string theory. However, there are other important types of
singularities that are not spacelike, and where current string
theory technology is unable to say what happens. My guess is that
some of them are acceptable and others are forbidden. But it
remains to be explained which is which and how this works.

\item {\it Understand and classify time-varying solutions.}
Only within the past couple of years have people built up the
courage to try to construct and analyze time-dependent solutions
to string theory. To start with, one goal is to construct examples
that can be analyzed in detail, and that do not lead to
pathologies. This seems to be very hard to achieve. This subject
seems to be badly in need of a breakthrough.

\item {\it Figure out which time-varying solution describes
the evolution of our Universe and understand whether it is special
or just an environmental accident.} If we had a complete list of
consistent time-dependent solutions, then we would face the same
sort of question we asked earlier in the particle physics context.
What is the principle by which a particular one is selected? How
much of the observed large-scale structure of the Universe can be
deduced from first principles? Was there a pre-big-bang era and
how did the Universe begin?
\end{itemize}

There is one last item that is of a somewhat different character
from the preceding ones, but certainly deserves to be included. We
need to
\begin{itemize}
\item {\it Develop the mathematical tools and concepts required
to solve all of the preceding problems.} String theory is up
against the frontiers of most major branches of mathematics. Given
our experience to date, there is little doubt that future
developments in string theory will utilize many mathematical tools
and concepts that do not currently exist. The need for cutting
edge mathematics is promoting a very healthy relationship between
large segments of the string theory and mathematics communities.
Such relationships were sadly lacking throughout a large part of
the twentieth century, and it is pleasing to see them blossoming
now.

\end{itemize}

\vspace{0.3cm}
{\bf Acknowledgements}.
This work was supported in part by the U.S. Dept. of Energy under
Grant No. DE-FG03-92-ER40701.


\begin{thereferences}{}

\bibitem{Aharony:2000}
Aharony, O., Gubser, S. S., Maldacena, J., Ooguri, H., \& Oz, Y.
2000,
Phys.\ Rep.\  {323}, 183

\bibitem{Banks:1997}
Banks, T., Fischler, W., Shenker, S. H., \& Susskind, L. 1997,
Phys.\ Rev.\ D, 55, 5112

\bibitem{Berenstein:2002}
Berenstein, D., Maldacena, J., \& Nastase, H. 2002,
JHEP, 0204, 013

\bibitem{Blau:2002}
Blau, M., Figueroa-O'Farrill, J., Hull, C., \& Papadopoulos, G.  2002,
JHEP, {0201}, 047

\bibitem{Brandenberger:1989}
Brandenberger, R. H., \& Vafa, C. 1989,
Nucl.\ Phys.\ B, {316}, 391

\bibitem{Brink:1977}
Brink, L., Schwarz, J. H., \& Scherk, J. 1977,
Nucl. Phys. B, 121, 77

\bibitem{Candelas:1985}
Candelas, P., Horowitz, G. T., Strominger, A., \& Witten, E. 1985,
Nucl.\ Phys.\ B, {258}, 46

\bibitem{Connes:1998}
Connes, A., Douglas, M. R., \& Schwarz, A. 1998,
JHEP, {9802}, 003

\bibitem{Cremmer:1978}
Cremmer, E., Julia, B., \& Scherk, J. 1978,
Phys.\ Lett.\ B, {76}, 409

\bibitem{Gervais:1971}
Gervais, J. L., \& Sakita, B. 1971,
Nucl. Phys. B, 34, 632

\bibitem{Giveon:1994}
Giveon, A., Porrati, M., \& Rabinovici, E. 1994,
Phys.\ Rep.,  {244}, 77

\bibitem{Gliozzi:1976}
Gliozzi, F., Scherk, J., \& Olive, D. 1976,
Phys. Lett. B, 65, 282

\bibitem{Gliozzi:1977}
------. 1977,
Nucl. Phys. B, 122, 253

\bibitem{Green:1983a}
Green, M. B., \& Schwarz, J. H. 1983,
Nucl.\ Phys.\ B, {218}, 43

\bibitem{Green:1984}
------. 1984,
Phys.\ Lett.\ B, {149}, 117

\bibitem{Green:1983b}
Green, M. B., Schwarz, J. H., \& Brink, L. 1983,
Nucl.\ Phys.\ B, {219}, 437

\bibitem{Green:1987}
Green, M. B., Schwarz, J. H., \& Witten, E. 1987, {Superstring
Theory} (Cambridge: Cambridge Univ. Press)

\bibitem{Greene:1990}
Greene, B. R., \& Plesser, M. R. 1990,
Nucl.\ Phys.\ B, {338}, 15

\bibitem{Gross:1985}
Gross, D. J., Harvey, J. A., Martinec, E. J., \& Rohm, R. 1985,
Phys.\ Rev.\ Lett.,  {54}, 502

\bibitem{Gubser:1998}
Gubser, S. S., Klebanov, I. R., \& Polyakov, A. M. 1998,
Phys.\ Lett.\ B, {428}, 105

\bibitem{Hawking:1976}
Hawking, S. W. 1976,
Phys.\ Rev.\ D, {14}, 2460

\bibitem{He:2002}
He, Y. H., Schwarz, J. H., Spradlin, M., \& Volovich, A. 2002,
preprint (hep-th/0211198)

\bibitem{Horava:1996}
Horava, P., \& Witten, E. 1996,
Nucl.\ Phys.\ B, {460}, 506

\bibitem{Hull:1994}
Hull, C. M., \& Townsend, P. K. 1995,
Nucl.\ Phys.\ B, {438}, 109

\bibitem{Maldacena:1998}
Maldacena, J. M. 1998,
Adv.\ Theor.\ Math.\ Phys.,  {2}, 231

\bibitem{Metsaev:2002}
Metsaev, R. R. 2002,
Nucl.\ Phys.\ B, {625}, 70

\bibitem{Nambu:1970}
Nambu, Y. 1970,
in Proc. Intern. Conf. on Symmetries and Quark Models, ed. R.
Chand (New York: Gordon and Breach), 269

\bibitem{Neveu:1971}
Neveu, A., \& Schwarz, J. H. 1971,
Nucl.\ Phys.\ B, 31, 86

\bibitem{Nielsen:1970}
Nielsen, H. B. 1970, unpublished

\bibitem{Obers:1999}
Obers, N. A., \& Pioline, B. 1999,
Phys.\ Rep.,  {318}, 113

\bibitem{Pankiewicz:2002}
Pankiewicz, A., \& Stefanski, B. 2003,
Nucl.\ Phys.\ B, 657, 79

\bibitem{Penrose:1976}
Penrose, R. 1976, in Differential Geometry and Relativity
(Dordrecht: Reidel), 271

\bibitem{Polchinski:1995}
Polchinski, J. 1995,
Phys.\ Rev.\ Lett.,  {75}, 4724

\bibitem{Polchinski:1998}
------. 1998, {String Theory} (Cambridge: Cambridge Univ. Press)

\bibitem{Ramond:1971}
Ramond, P. 1971,
Phys.\ Rev.\  D, 3, 2415

\bibitem{Santambrogio:2002}
Santambrogio, A., \& Zanon, D. 2002,
Phys.\ Lett.\ B, {545}, 425

\bibitem{Scherk:1974}
Scherk, J., \& Schwarz, J. H. 1974,
Nucl. Phys. B, 81, 118

\bibitem{Schwarz:1983}
Schwarz, J. H. 1983,
Nucl.\ Phys.\ B, {226}, 269

\bibitem{Seiberg:1999}
Seiberg, N., \& Witten, E. 1999,
JHEP, {9909}, 032

\bibitem{Spradlin:2002a}
Spradlin, M., \& Volovich, A. 2002,
Phys.\ Rev.\ D, {66}, 086004

\bibitem{Spradlin:2002b}
------. 2003,
JHEP, 0301, 036

\bibitem{Strominger:1996}
Strominger, A., \& Vafa, C. 1996,
Phys.\ Lett.\ B, {379}, 99

\bibitem{Susskind:1970}
Susskind, L. 1970,
Nuovo Cim., {69A}, 457

\bibitem{Townsend:1995}
Townsend, P. K. 1995,
Phys.\ Lett.\ B, {350}, 184

\bibitem{Veneziano:1968}
Veneziano, G. 1968,
Nuovo Cim., {57A}, 190

\bibitem{Wess:1974}
Wess, J., \& Zumino, B. 1974,
Nucl. Phys. B, 70, 39

\bibitem{Witten:1995}
Witten, E. 1995,
Nucl.\ Phys.\ B, {443}, 85

\bibitem{Witten:1996}
------. 1996,
Nucl.\ Phys.\ B, {460}, 335

\bibitem{Witten:1998}
------. 1998,
Adv.\ Theor.\ Math.\ Phys.,  {2}, 253

\bibitem{Yoneya:1974}
Yoneya, T. 1974,
Prog. Theor. Phys., {51}, 1907

\end{thereferences}
\end{document}